\documentclass[aps,tightenlines,longbibliography,notitlepage,nofootinbib]{revtex4-2}

\usepackage{amsmath}
\usepackage{amssymb}
\usepackage{bbm}
\usepackage{simplewick}
\usepackage{mathtools}
\usepackage[normalem]{ulem}
\usepackage{dsfont}
\usepackage{mathrsfs}
\usepackage[makeroom]{cancel}
\usepackage{graphicx}
\usepackage{bm}

\DeclareMathOperator\im{Im}

\def\rhat{{\boldsymbol{\hat{r}}}}
\def\khat{{\boldsymbol{\hat{k}}}}

\DeclareMathOperator\daw{daw}
\DeclareMathOperator\erf{erf}

\def\0{\boldsymbol{0}}
\def\RR{\mathbbm{R}}

\def\BD#1{{\boldsymbol{{#1}}}}

\def\ii{\mathrm{i}}
\def\bnabla{{\boldsymbol\nabla}}
\def\OM#1{\omega_{#1}}

\def\B#1{\left(#1\right)}
\def\BB#1{\left[#1\right]}

\def\la{\langle}
\def\ra{\rangle}
\def\lara#1{\la#1\ra}

\def\be{\begin{equation}}
\def\ee{\end{equation}}

\def\hc{\text{h.c.}}
\def\cc{\text{c.c.}}

\def\for{\ \text{for} \ }

\usepackage[ddmmyyyy]{datetime}
\makeatletter
\def\Dated@name{}
\makeatother

\def\pref{0.5}
\def\KS{{\BD{k}\sigma}}
\def\jmat{\boldsymbol{j}_\text{mat}}

\begin{document}
\title{Expanding quantum magnetic  field}
\author{Bogdan Damski}
\affiliation{\text{Jagiellonian University,
Faculty of Physics, Astronomy and Applied Computer Science,}
\\ {\L}ojasiewicza 11, 30-348 Krak\'ow, Poland}
\begin{abstract}
We develop the  quantum theory of
the causal formation of a long-range magnetic field
generated by an  external current that is instantaneously switched 
on and subsequently kept constant in time.
The resulting non-equilibrium  quantum state, describing the 
expanding magnetic field,  is obtained exactly
and compared with the corresponding quantum magnetostatic state.
In contrast to the magnetostatic case,
 the expanding solution exhibits 
a propagating shockwave-like front separating regions where 
the magnetic field has already been  formed from those
 that remain causally disconnected from the source.
We show that although the
expanding field locally approaches the magnetostatic field  behind the shockwave-like
front, the associated quantum systems remain distinct  at all times. 
In particular, we obtain manifestly different results for the 
energy, photon number, and their fluctuations in expanding 
and magnetostatic field configurations.
Our results are first derived for a general external 
current and then illustrated with a specific example.

\end{abstract}
\maketitle

\section{Introduction}

A magnetostatic field is  introduced  in the 
following standard way \cite{JacksonBook,GriffithsBook}. 
One assumes that an electric charge density 
is constant in time, specifies a time-independent external current,
and then  solves the  Maxwell's equations. This way one obtains
a static  field configuration, where the  
magnetic field algebraically decays far away from the region, where the 
 current is localized.
Since this decay extends, 
at any given instant, over arbitrarily large distances, 
one may wonder how the field could have propagated that far. 
It turns out that this is a consequence of 
assuming the past-eternal presence of the current, 
a premise that is clearly 
invalid when discussing magnetic fields generated in a laboratory 
or elsewhere in the Universe over finite time scales.

This motivates the study of a more physically plausible scenario, 
where the external current appears at some instant, say $t=0$,
and then does not change in time.
This scenario (i)  can be formulated so as to preserve charge conservation,
(ii) represents the simplest   modification of the static current problem,
and (iii) facilitates  analytical  comparison of the obtained results 
to the static ones.
We discuss such a scenario in this work assuming for simplicity
that an electric charge density vanishes (just as in 
a  current-in-a-loop-wire setup).

Under such circumstances,  
one may expect on general grounds that 
the magnetic field will be established, in  otherwise empty space,
at the speed of light.
In particular, in the extreme limit 
of the point-like support of the current, one may speculate 
that there should  be a ``static'' magnetic field 
in the ball of radius $t$, an electric field localized  on the
surface of such a ball, and 
neither magnetic nor electric
field outside this ball (the ball is
assumed to be centered at the point, where the current is localized).
Qualitatively, the same behavior is expected in a setup with an extended 
but  localized current, except  there shall be 
a smooth crossover to the field-free  region. 
Either way, it is expected that 
there will be an expanding field in the system accompanied by a
shockwave-like front.

The issue of the dynamical  creation of a long-range field was considered
in \cite{BermanAmJPhys2008}
in the electrostatic   context
within the classical   Maxwell theory.
As discussed in \cite{BermanAmJPhys2008}, this process, 
due to electric charge conservation, 
requires appearance of a  current transferring a
charge all the way from spatial infinity, which is 
questionable.
 To overcome this problem, 
 it is assumed in \cite{BermanAmJPhys2008} that  two 
charges of the same magnitude but opposite sign are created 
at some finite separation, and then the limit is taken 
in which the separation (magnitude) of the charges goes to zero (infinity)
such that  the dipole moment is fixed.
This procedure produces cancellations that make the resulting current effectively localized at a point.
However, since all physical electric dipoles we know consist
of charges separated by a non-zero  distance and having a finite 
magnitude, 
 the above approach does not correspond to any real physical situation.
We find it important to  emphasize  that charge conservation 
does not lead to  any such complications in the studies discussed 
in our work.

The questions now are the following. Can  a  magnetostatic
field be regarded as the $t\to\infty$ version of 
its expanding counterpart? What theoretical framework should 
be employed for discussing this  issue?

Regarding the latter question, we observe that the dynamics 
of an electromagnetic field, resulting from  time dependence
of an external current, is traditionally discussed within the
classical framework in the context of electromagnetic
wave generation \cite{Landau,JacksonBook,GriffithsBook}. 
In this approach, one obtains causal solutions to the Maxwell's equations
under suitable initial conditions. Alternatively, one may 
use Jefimenko's equations \cite{Jefimenko2004}, which 
directly relate the electromagnetic field 
in the discussed setting to the external current and its time 
derivative (both evaluated at retarded times).
We shall not use the classical  theory in this work.
Instead, we work within the quantum framework solving the Schr\"odinger equation.
This yields a wavefunction allowing  for the  quantitative 
comparison of the static and expanding magnetic fields. It also allows 
the critical comparison of photon numbers and fluctuations of
various quantities  characterizing these  fields,
which is  beyond the realm of the classical framework.

Regarding the former question, we argue below that  quantum 
states representing the static  and expanding magnetic  fields
fundamentally differ no matter
how far in the past the current generating the expanding magnetic field was
established.

The outline of this paper is the following.
Sec. \ref{Basics_sec} formulates the observables of interest within
the Fock-space formalism and clarifies the shockwave terminology.
Sec. \ref{Static_sec} discusses 
the coherent state structure of a 
quantum state representing a
magnetostatic field.
Sec. \ref{Time_sec} provides formulas describing  the dynamics 
induced by an instantaneously switched on external current that 
subsequently  evolves arbitrarily.
Properties of the system after the sudden onset of a 
static external current are studied in Sec. \ref{Problem_sec}.
Technical insights  pertaining to the evaluation of integrals are 
given in Sec.~\ref{Sum_sec}.
The discussion in Secs. \ref{Static_sec} and \ref{Problem_sec}
is quite general as it is carried out without assuming any explicit 
spatial structure of an external  current. 
Specific illustrations of these findings are presented 
in Secs. \ref{Static_sec2} and \ref{Expanding_sec}. 
Orders of magnitude of the quantities discussed 
in  Secs. \ref{Static_sec2} and \ref{Expanding_sec}
are analyzed   in Sec. \ref{Orders_sec}.
The summary of our work is provided in Sec. \ref{Summary_sec}.

\section{Basics}
\label{Basics_sec}

We   work with the system, 
where the charge density vanishes everywhere in space. 
Thereby, in the Coulomb gauge that we employ, the electromagnetic 
field is  described by the transverse vector field operator $\BD{A}$.
This  vector field is  coupled to 
the external current $\BD{j}_\text{mat}$.

The  normal-ordered Schr\"odinger-picture  Hamiltonian
describing this  system is \cite{Greiner}
\begin{subequations}
\begin{align}
& H(0;t)= H_0 + H_1(0;t),\\
& H_0= \frac{1}{2}\int d^3r:\BD{E}(0,\BD{r})\cdot\BD{E}(0,\BD{r}) +
\BD{B}(0,\BD{r})\cdot\BD{B}(0,\BD{r}):,\\
& H_1(0;t)= -\int d^3r \jmat(t,\BD{r})\cdot \BD{A}(0,\BD{r}),
\end{align}
\label{HHH}%
\end{subequations}
where $H_0$ ($H_1$) stands for the  free-field (interaction)
Hamiltonian while the electric and magnetic field operators
read  $\BD{E}=-\partial_t \BD{A}$ and 
$\BD{B}=\bnabla\times\BD{A}$; see 
Appendix~\ref{Conventions_app}
for our conventions.
Three  remarks are in order now.

First, the first argument of  $H_1$ 
refers to  time
dependence of the {\it operator}  $\BD{A}$ involved in the definition
of $H_1$. The second argument of $H_1$ refers to time dependence
of $\jmat$. 
The same  remarks explain the arguments of $H$. We stress that 
time dependence of $\jmat$ is unaffected by choosing the 
Schr\"odinger picture  because $\jmat$ is 
a {\it classical}  current.
Moreover, the external current  will be chosen so as to be conserved. This
requirement, in the system where the charge density vanishes everywhere in
space, implies 
\be
\bnabla\cdot\jmat=0.
\label{nabjmat}
\ee

Second, Schr\"odinger-picture $\BD{A}(0,\BD{r})$ appearing above is given by \cite{Greiner}
\begin{align}
\label{A0r}
&\BD{A}(0,\BD{r})= \int
\frac{d^3k}{(2\pi)^{3/2}}
\frac{1}{\sqrt{2\OM{k}}}
\sum_\sigma
\BD{\eta}(\BD{k},\sigma)a_\KS\exp(\ii\BD{k}\cdot\BD{r})
+\text{h.c.},\\
&
 \lbrack a_{\BD{k}\sigma},a^\dag_{\BD{k}'\sigma'}\rbrack=\delta_{\sigma\sigma'}\delta(\BD{k}-\BD{k}'), 
 \ \lbrack a_{\BD{k}\sigma},a_{\BD{k}'\sigma'}\rbrack=0,\\
 & \BD{\eta}(\BD{k},\sigma)\in\RR^3, \
\BD{\eta}(\BD{k},\sigma)\cdot\BD{\eta}(\BD{k},\sigma')=\delta_{\sigma\sigma'}, \
\BD{\eta}(\BD{k},\sigma)\cdot\BD{k}=0,
\end{align}
where $\OM{k}=|\BD{k}|$ while $\sum_\sigma\equiv\sum_{\sigma=1}^2$
denotes a summation over   photon polarizations.
Moreover, the Schr\"odinger-picture electric field operator,
which is  the negative of the momentum canonically
conjugate to (\ref{A0r}),  reads    \cite{Greiner}
\be
\label{ewdewk}
 \BD{E}(0,\BD{r})
 =\ii \int\frac{d^3k}{(2\pi)^{3/2}}
\sqrt{\frac{\OM{k}}{2}}
\sum_\sigma
\BD{\eta}(\BD{k},\sigma)a_\KS\exp(\ii\BD{k}\cdot\BD{r})
+\hc
\ee
Its magnetic  counterpart, $\BD{B}(0,\BD{r})$, 
can be obtained from (\ref{A0r}). 

Third, using the above  expressions, one may  verify
that \cite{Greiner}
\be
H_0=\int d^3k \OM{k}\sum_{\sigma} a^\dag_\KS a_\KS.
\label{0H0}
\ee
The vacuum state $|0\ra$ is the  ground state of $H_0$. It 
 is  annihilated by all  operators $a_\KS$.

Finally, regarding terminology, we will use the expression
{\it shockwave-like} when discussing the fields in our system.
This expression is intended to denote a propagating feature that varies rapidly in space.
The fields considered in this work  satisfy linear
differential equations, and so  one does not expect them to encode genuine 
shockwaves 
(e.g. the shockwave-like features analyzed  in this work do not undergo self-steepening dynamics).
It is  worth emphasizing that the question of the  existence of quantum shockwaves is itself quite
delicate.
For example,  quantum shockwaves arise in  cold 
non-interacting Fermi gases, despite the linearity of an underlying many-body quantum  problem, 
when one considers the dynamics of a single-particle density \cite{BDJPB2004}.
A somewhat similar situation is found  in  interacting cold  Bose gases 
(see e.g. \cite{QuantumShock2020,Olshanii2021} for recent contributions).

\section{Time independent  external current: general results}
\label{Static_sec}

We show below that the ground state of the system in the presence of a static 
external current is represented by a coherent state. The calculations leading to
this conclusion are standard (see e.g. \cite{PandA} for their electrostatic counterpart),
whereas the discussion of the properties of the ground state is 
less conventional (see e.g. the remarks about the distribution of the 
energy between free-field and interaction Hamiltonians).
Coherent states, first discussed by Schr\"odinger a hundred years ago, are well known in
various branches of physics.\footnote{An English translation of the 
relevant  Schr\"odinger's  paper can be found
on pp.~41--44 of \cite{Schrodinger}.} It is worth noting that they play a 
central role in quantum optics, where they are used to model laser light, 
electromagnetic fields inside optical cavities, 
the local oscillator in homodyne detection,  etc. (see e.g. \cite{Gerry,Scully}).
While the techniques used to describe such applications of coherent states 
are also applicable to our studies, the emphasis in our  work is on 
long-distance features, which are  of secondary interest in the quantum optics context.
A different situation is found  in scattering problems
in quantum electrodynamics, where bare electron and positron states 
are dressed by coherent states of long-wavelength photons 
in the Chung \cite{Chung1965} and Kulish-Faddeev \cite{KulishFaddeev}
approaches. 
These  dressings  were  introduced  in order to remove infrared (IR) 
singularities from scattering amplitudes. 
In-depth studies of these  constructions and their implications 
remain ongoing
(see e.g. \cite{StromingerPRD2017,CarneyPRD2018,TomarasPRD2020,FurugoriPRD2021}).
The similarity between these considerations and our work is primarily technical
(e.g. because we work with an external current).

To proceed,  we  put time-independent current  $\jmat(\BD{r})$  into (\ref{HHH}) getting  
the Hamiltonian, which can be written in  the following diagonal form 
\be
H=\sum_\KS \OM{k}   \tilde{b}^\dag_\KS \tilde{b}_\KS
-\sum_\KS \OM{k} |\lambda_\KS|^2,
\label{Hdiag}
\ee
where we have switched from the continuous quantization model to the 
quantization-in-a-periodic-box one via the following standard identities
($V$ is the volume in which the theory is quantized)
\begin{align}     
\label{qbox1}
&\int \frac{d^3k}{(2\pi)^{3/2}}a_\KS \mapsto \sum_\BD{k} \frac{b_\KS}{\sqrt{V}},\\
\label{qbox2}
&\int d^3k a^\dag_\KS a_\KS \mapsto \sum_\BD{k} b^\dag_\KS b_\KS,\\
&[b_\KS,b^\dag_{\BD{k}'\sigma'}]= \delta_{\sigma\sigma'}  \delta_{\BD{k}\BD{k}'},
\  [b_\KS,b_{\BD{k}'\sigma'}]=0.
\end{align}
Note that there are no time arguments in  $H$  due to the static character of the
assumed current.\footnote{The same remark  applies to $H_1$ in (\ref{h0h1b}) and
(\ref{H0-2}).}
In the above formulas,  
\begin{align}
\label{rfnj}
&\tilde{b}_\KS= b_\KS -\lambda_\KS,\\
&\lambda_\KS=\frac{\BD{J}_\text{mat}(\BD{k})\cdot\BD{\eta}(\BD{k},\sigma)}{\sqrt{2V}\OM{k}^{3/2}},
\label{lLKS}\\
&\BD{J}_\text{mat}(\BD{k}) =  \int d^3r \jmat(\BD{r})\exp(-\ii\BD{k}\cdot\BD{r}),\\
&
[\tilde{b}_\KS,\tilde{b}^\dag_{\BD{k}'\sigma'}]=\delta_{\sigma\sigma'} \delta_{\BD{k}\BD{k}'}, \ 
[\tilde{b}_\KS,\tilde{b}_{\BD{k}'\sigma'}]=0.
\end{align}

We introduce the state $|\tilde{0}\ra$, the ground state of $H$,
which  is annihilated by all operators $\tilde{b}_\KS$. It is a coherent state 
\begin{align}
&|\tilde{0}\ra= \bigotimes_\KS |\lambda_\KS\ra, \\
\label{coooh}
&b_\KS |\lambda_\KS\ra = \lambda_\KS |\lambda_\KS\ra,\\
&|\lambda_\KS\ra= D(\lambda_\KS) |0\ra,\\
&D(\lambda_\KS)=\exp(\lambda_\KS b^\dag_\KS - \lambda^*_\KS b_\KS),
\end{align}
where $D$ is called the shift operator due to (\ref{Ddef}); see e.g.  
\cite{Gerry,Scully} for a discussion of basic properties of coherent states.
The   eigenenergy of $|\tilde{0}\ra$ is 
\be
\tilde{\cal E}=-\sum_\KS \OM{k}  |\lambda_\KS|^2,
\label{tildeE}
\ee
and so we introduce the Schr\"odinger-picture wavefunction 
\be
 |\tilde{\psi}(t)\ra=\exp(-\ii\tilde{\cal E} t)|\tilde{0}\ra.
 \label{psiTt}
\ee
To set the ground for the coming discussion,
we note the obvious fact that the standard deviation  of the energy in the 
state $|\tilde{0}\ra$,
\be
\tilde{\sigma}_{\cal E}=\sqrt{ \lara{\tilde{0}|H^2 |\tilde{0} }
-\lara{\tilde{0}|H |\tilde{0}}^2      },
\ee
vanishes.

As far as the distribution of the energy 
between free-field and interaction
 Hamiltonians is concerned, we have the following remarks.
 Using the above expressions, it is easy to find that 
\begin{align}
\label{h0h1a}
&\lara{\tilde{\psi}(t)|H_0|\tilde{\psi}(t)}= -\tilde{\cal E},\\
&\lara{\tilde{\psi}(t)|H_1|\tilde{\psi}(t)} =2 \tilde{\cal E},
\label{h0h1b}
\end{align}
which can be also  obtained from
the considerations based on the  so-called Hellmann-Feynman theorem.
Thereby, 
\be
\lara{\tilde{\psi}(t)|H_1|\tilde{\psi}(t)} =
-2\lara{\tilde{\psi}(t)|H_0|\tilde{\psi}(t)},
\label{H0-2}
\ee
which is reminiscent of the virial theorem. We 
mention in passing that such a relation does not hold for 
excited   eigenstates of Hamiltonian (\ref{Hdiag}); see 
Appendix~\ref{App_Virial}.

Finally, we introduce the  photon number operator
\be
N=\sum_\KS b_\KS^\dag b_\KS.
\ee
We  use it to compute the number of photons and its 
standard deviation in the state $|\tilde{\psi}(t)\ra$
\begin{align}
\label{Nph}
&\tilde{\cal N}=\lara{\tilde{\psi}(t)|N|\tilde{\psi}(t)}
=\sum_\KS |\lambda_\KS|^2,\\
&\tilde{\sigma}_{\cal N}=\sqrt{\lara{\tilde{\psi}(t)|N^2|\tilde{\psi}(t)}
-
\lara{\tilde{\psi}(t)|N|\tilde{\psi}(t)}^2}=\sqrt{\tilde{\cal N}}.
\label{VarN}
\end{align}
The discussion of the static problem is concluded with the expression for the 
expectation value of the vector field operator 
\be
\lara{\tilde{\psi}(t)|\BD{A}(0,\BD{r})|\tilde{\psi}(t)}=
\sum_\KS \frac{\BD{\eta}(\BD{k},\sigma)}{\sqrt{2V\OM{k}}}\lambda_\KS
\exp(\ii\BD{k}\cdot\BD{r})+\cc,
\label{VVVstat}
\ee
whose curl  yields
$\lara{\tilde{\psi}(t)|\BD{B}(0,\BD{r})|\tilde{\psi}(t)}$.
As anticipated on general grounds,  
$\lara{\tilde{\psi}(t)|\BD{E}(0,\BD{r})|\tilde{\psi}(t)}=\0$.

\section{Time dependent  external current: general results}
\label{Time_sec}

The problem of coherent-state generation by an external current
is briefly discussed in various  textbooks \cite{Gerry,Scully,PandA}. 
We adopt the formalism presented there  for our purposes and
extend it in a way  that is explained  below (\ref{cohI}).
The goal here is to derive expressions describing the quantum dynamics 
induced by a time-dependent external current. 
In this case, the Schr\"odinger-picture Hamiltonian is explicitly
time dependent and the problem is further complicated by the fact
that the Hamiltonians at different times do not commute.
Still, the problem can be elegantly solved in the interaction picture and the 
resulting expressions can be readily expressed in the Schr\"odinger and
Heisenberg pictures. 
These results will be obtained by performing the
calculations in the quantization-in-a-periodic-box scheme.

We assume that the system is initially
in the vacuum state,
\be
|\psi(0)\ra=|0\ra,
\label{psi0}
\ee
which makes sense when an external current vanishes for $t<0$.
In what follows, we  assume that $t\ge0$.

The interaction-picture vector field  reads 
\be
\label{AIII}
\BD{A}_I(t,\BD{r})= \exp(\ii H_0 t) \BD{A}(0,\BD{r})\exp(-\ii H_0 t)=
\sum_\KS
\frac{\BD{\eta}(\BD{k},\sigma)}{\sqrt{2V\OM{k}}}
b_\KS \exp(-\ii\OM{k}t+\ii\BD{k}\cdot\BD{r})
+\text{h.c.},
\ee
whereas the interaction-picture version of $H_1(0;t)$ is 
\be
H_1^I(t)=\exp(\ii H_0 t) H_1(0;t)  \exp(-\ii H_0 t)
=-\sum_\KS \frac{\BD{J}^*_\text{mat}(t,\BD{k})\cdot\BD{\eta}(\BD{k},\sigma)}{\sqrt{2V\OM{k}}}
b_\KS \exp(-\ii\OM{k}t)+\hc,
\ee
where 
$\BD{J}_\text{mat}(t,\BD{k})=\int d^3r \jmat(t,\BD{r})\exp(-\ii\BD{k}\cdot\BD{r})$. 
Note that  there is no need to  introduce the second time argument in $H_1^I$ 
and that the label $I$ denotes  interaction-picture quantities.

The interaction-picture wavefunction is 
\be
|\psi_I(t)\ra = U_I(t,0)|\psi(0)\ra, 
\, U_I(t,0)=\mathbbm{T}\exp\B{-\ii \int_0^t dt' H_1^I(t')},
\ee
where $\mathbbm{T}$ is the time-ordering operator.
To simplify this  expression, we introduce $t_j=j dt$ and observe that 
\be
U_I(t,0)= \lim_{dt\to0^+} \prod_{j=0}^{t/dt-1} \exp(-\ii H^I_1(t_j)dt)
\equiv \lim_{dt\to0^+} \exp(-\ii H^I_1(t_{t/dt-1})dt)  \cdots  \exp(-\ii H^I_1(0)dt),
\ee
which can be written as 
\begin{align}
& U_I(t,0)= \prod_\KS\B{ \lim_{dt\to0^+}\prod_{j=0}^{t/dt-1} D\B{\beta_\KS(t_j)dt}},\\
& \beta_\KS(t)=  \frac{\ii \exp(\ii\OM{k}t) }{\sqrt{2V\OM{k}}}
\BD{J}_\text{mat}(t,\BD{k})\cdot\BD{\eta}(\BD{k},\sigma).
\label{betat}
\end{align}
Taking into account  that
\be
D(\gamma_\KS)D(\xi_\KS) =
\exp\B{\ii\text{Im}(\gamma_\KS \xi^*_\KS)} D(\gamma_\KS+\xi_\KS),
\ee
we have obtained
\begin{align}
\label{UIexplicit}
& U_I(t,0)=\prod_\KS  \exp\B{\ii\text{Im}\Phi_\KS(t)} D\B{\alpha_\KS(t)},\\
& \alpha_\KS(t)= \int_0^t dt' \beta_\KS(t'),\\
& \Phi_\KS(t)=\int_0^t dt'\beta_\KS(t')\int_0^{t'}
dt''\beta_\KS^*(t'').
\end{align}

Using (\ref{psi0}), we  arrive at the following  compact expression
\be
|\psi_I(t)\ra =
\bigotimes_\KS \exp\B{\ii\text{Im}\Phi_\KS(t)}  |\alpha_\KS(t)\ra,
\label{cohI}
\ee
where we would like to emphasize that the expression for the above phase factor is 
missing  in 
standard textbooks \cite{Gerry,Scully,PandA}. 
Useful information will be extracted out 
of this  phase factor in Sec. \ref{Fluctuations_sub}.
Equation (\ref{cohI}), along with the above results for interaction-picture operators,
finishes our discussion of interaction-picture expressions.
Their counterparts in other pictures are the following.

In the Heisenberg picture, the vector field  operator 
reads 
\be
\BD{A}(t,\BD{r})= U_I^\dag(t,0) \BD{A}_I(t,\BD{r}) U_I(t,0) = 
\BD{A}_I(t,\BD{r}) +
\sum_\KS \frac{  \BD{\eta}(\BD{k},\sigma)   }{\sqrt{2V\OM{k}}}
\B{\alpha_\KS(t)
\exp(-\ii\OM{k}t+\ii\BD{k}\cdot\BD{r})
+\cc},
\label{HeiI}
\ee
which can be shown via the following identity
\be
 D^\dag(\gamma_\KS) b_\KS D(\gamma_\KS) = b_\KS + \gamma_\KS.
 \label{Ddef}
\ee
In the Schr\"odinger picture, the  wavefunction is 
\be
|\psi(t)\ra= \exp(-\ii H_0 t)|\psi_I(t)\ra
=\bigotimes_\KS \exp\B{\ii\text{Im}\Phi_\KS(t)} 
|\alpha_\KS(t)\exp(-\ii\OM{k}t) \ra,
\label{SchSch}
\ee
which can be  verified using  the following standard
formula
\be
|\gamma_\KS\ra= \exp\B{- \frac{|\gamma_\KS|^2}{2}}
\sum_{n=0}^\infty \frac{\gamma_\KS^n}{\sqrt{n!}}|n_\KS\ra,
\ee
where $|n_\KS\ra=(b^\dag_\KS)^n|0_\KS\ra/\sqrt{n!}$ is the Fock state containing 
 $n$ photons with  momentum $\BD{k}$ and polarization $\sigma$
 ($b_\KS|0_\KS\ra=0$, $|0\ra=\bigotimes_\KS|0_\KS\ra$).

\section{Problem setup}
\label{Problem_sec}
We are now  interested in the  dynamics induced by the 
following external current 
\be
\jmat(t,\BD{r})=\jmat(\BD{r})\theta(t),
\label{ewfkmlk}
\ee
where $\theta(t)$ is the Heaviside step function. 
We again assume (\ref{psi0}) and, unless stated otherwise, 
provide  expressions relevant for  $t\ge0$.
For this  current, 
\begin{align}
\label{alfaG}
&\alpha_\KS(t)= \B{\exp(\ii\OM{k}t)-1}\lambda_\KS,\\
&\Phi_\KS(t)=\B{1+\ii \OM{k}t -\exp(\ii \OM{k}t)}|\lambda_\KS|^2
\label{PhiG}
\end{align}
characterize the time-evolved state $|\psi(t)\ra$. These 
parameters  are linked to the static  problem via $\lambda_\KS$ (\ref{lLKS}).
The question now is how we can quantify differences between
the states $|\tilde{\psi}(t)\ra$ and $|\psi(t)\ra$
representing the  static and expanding fields, 
respectively.\footnote{The expanding nature of the magnetic 
field encoded 
in the state $|\psi(t)\ra$ will be illustrated in Sec.~\ref{Expanding_sec}.
}
 This can be done by 
considering  the vector field in these states (Sec. \ref{Vector_sub}) and 
the overlap between them  (Sec. \ref{Overlap_sub})
as well as the energy   (Sec. \ref{Energy_sub}),
the energy fluctuations (Sec. \ref{Fluctuations_sub}), and the number 
of photons (Sec. \ref{Photon_sec}) in the state $|\psi(t)\ra$.
These issues  will be discussed below without assuming a specific 
form of $\jmat(\BD{r})$. 
We assume only that $\jmat(\BD{r})$ leads to convergent 
expressions for the discussed quantities 
and that the Riemann–Lebesgue lemma applies whenever invoked.
Finally, we observe that (\ref{ewfkmlk})
results in a sudden change of the Hamiltonian that is otherwise time
independent.
Thereby, adopting the terminology commonly employed in the field of non-equilibrium 
phase transitions \cite{DziarmagaAdvPhys,PolkovnikovRMP},
one may say that the discussed system undergoes a sudden quench.

\subsection{Vector field}
\label{Vector_sub}
We have 
\be
\lara{\psi(t)|\BD{A}(0,\BD{r})|\psi(t)}=
\sum_\KS
\frac{\BD{\eta}(\BD{k},\sigma)}{\sqrt{2V\OM{k}}}\alpha_\KS(t)
\exp(-\ii\OM{k}t+\ii\BD{k}\cdot\BD{r})+\cc,
\ee
which via  (\ref{alfaG}) can be written as
\be
\lara{\psi(t)|\BD{A}(0,\BD{r})|\psi(t)}=
\lara{\tilde{\psi}(t)|\BD{A}(0,\BD{r})|\tilde{\psi}(t)}
-\sum_\KS
\frac{\BD{\eta}(\BD{k},\sigma)}{\sqrt{2V\OM{k}}}\B{\lambda_\KS
\exp(-\ii\OM{k}t+\ii\BD{k}\cdot\BD{r})+\cc}.
\label{VVV}
\ee
The message delivered by this equation is clear when one considers 
a fixed point  in space, say $\BD{r}$, and the limit of
$t\to\infty$.
Then, one may use the  Riemann–Lebesgue lemma 
to specify the class of external currents for which the second
term  of (\ref{VVV}) vanishes. 
When this happens, one is left, asymptotically in time, 
with the static result at the point $\BD{r}$.
However, one should not conclude that this observation implies that the 
static field is established throughout all space 
because the above reasoning is carried out under the tacit assumption that 
$|\BD{r}|\ll t$, which is implicit in the phrase ``a fixed point in space''.
In fact,  fundamental differences between the static and expanding fields 
will be identified below.
Finally, we observe that the causal formation  of the vector field is not apparent from (\ref{VVV}).
This will be illustrated in Sec. \ref{Expanding_sec} through the specific example.

\subsection{Overlap}
\label{Overlap_sub}

The overlap between Schr\"odinger-picture
states representing static and expanding fields reads
\be
\lara{\tilde{\psi}(t)|\psi(t)} = \exp(\ii\tilde{\cal E}t)
\prod_\KS \exp\B{\ii\text{Im}\Phi_\KS(t)}
 \lara{\lambda_\KS|\alpha_\KS(t)\exp(-\ii\OM{k}t)}.
\label{FFf}
\ee
Using (\ref{alfaG}), we see that 
\begin{align}
\label{efwlk}
&\alpha_{\BD{k}\sigma}(t)\exp(-\ii\OM{k}t) = \lambda_{\BD{k}\sigma}
-\lambda_{\BD{k}\sigma}\exp(-\ii\OM{k}t),
\end{align}
which leads to two comments.

First, $\alpha_{\BD{k}\sigma}(t)\exp(-\ii\OM{k}t)$ 
differs from $\lambda_{\BD{k}\sigma}$ for all times: the state 
representing the static 
 field is never obtained  from its expanding counterpart,
not even  in the limit 
of $t\to\infty$. In fact, as far as (\ref{efwlk}) is concerned, 
this limit  does not exist.

Second, we note that
the Hamiltonian is time independent
in the discussed  problem,
 except at $t=0$ when it 
changes instantaneously. 
Thereby, $|\psi(t)\ra$ undergoes free evolution   for $t>0$
\be
|\psi(t)\ra = \tilde{a}  \exp(-\ii\tilde{\cal E}t) |\tilde{0}\ra + 
\sum_\alpha \tilde{a}_\alpha    \exp(-\ii\tilde{\cal E}_\alpha t) |\tilde{\alpha}\ra,
\label{dewklm}
\ee
where $|\tilde{\alpha}\ra$ are excited eigenstates of $H(0;t)=H(0;0^+)=H$
to eigenvalues $\tilde{\cal E}_\alpha$ and 
 amplitudes, $\tilde{a}$ and $\tilde{a}_\alpha$, are 
 time independent; see (\ref{Hdiag}) for $H$. Given that
$\lara{\tilde{0}|\tilde{\alpha}}=0$,
we gather  from the above formula  that $\lara{\tilde{\psi}(t)|\psi(t)}$ is
also time independent.
In fact, a direct calculation shows that 
\be
\lara{\tilde{\psi}(t)|\psi(t)} = 
\exp\B{ -\frac{\tilde{\cal N}}{2}  }. 
\label{overLAP}
\ee
To obtain this result, 
we have substituted (\ref{tildeE}), (\ref{alfaG}), and (\ref{PhiG}) into
(\ref{FFf}),   and  used
\be
\label{alfabeta}
\lara{\gamma_\KS|\xi_\KS} =
\exp\B{-\frac{|\gamma_\KS|^2}{2} -   \frac{|\xi_\KS|^2}{2}
+\gamma^*_\KS \xi_\KS}.
\ee
Two remarks are in order now.

First, the same result for (\ref{overLAP}) is obtained from $\lara{\tilde{0}|0}$. 
Therefore, (\ref{overLAP}) can be interpreted  as the overlap between two ground 
states of the same physical system corresponding to different values of the 
Hamiltonian parameters (i.e. vanishing or non-vanishing external current). 
From this perspective, (\ref{overLAP}) is similar to the ground-state 
fidelity known from the theory of equilibrium phase transitions 
\cite{Gu2010}.
Second, we note that (\ref{overLAP}) is necessarily smaller than one. 
It quantifies the difference between the states representing the   static and expanding fields.

\subsection{Energy}
\label{Energy_sub}

A simple calculation yields
\begin{align}
\label{H00000}
& \lara{\psi(t)|H_0|\psi(t)}= \sum_\KS \OM{k}|\alpha_\KS(t)|^2, \\
& \lara{\psi(t)|H_1(0;t)|\psi(t)}= -\ii\sum_\KS\beta^*_\KS(t)\alpha_\KS(t)+\cc
\end{align}
Using (\ref{alfaG}) and $\beta_\KS(t)=d\alpha_\KS(t)/dt$, we find  
\be
\lara{\psi(t)|H_1(0;t)|\psi(t)}
=-\lara{\psi(t)|H_0|\psi(t)}.
\label{H0H1}
\ee
This results in  two observations.

First, the relation between the expectation values of the 
free-field and interaction  Hamiltonians 
fundamentally differs between the static and expanding problems;
see (\ref{H0-2}). Moreover,  the energy of the expanding
 field vanishes
\be
{\cal E} = \lara{\psi(t)|H(0;t)|\psi(t)}=0.
\label{Hall}
\ee
This   null result 
is in stark contrast to the energy of the static  field configuration, 
which is necessarily negative (\ref{tildeE}).
This  can be explained as follows.
The Hamiltonian discontinuously changes  at $t=0$
but it has zero vacuum expectation value both 
before and after the change. The quantum state of the 
system continuously evolves from $t=0$ according to the 
Schr\"odinger equation. Thereby, $|\psi(0^+)\ra= |\psi(0)\ra=|0\ra$ 
and so the energy at $t=0^+$ vanishes. 
Given the fact that the Hamiltonian is time independent for
$t>0$, the energy is conserved and so it vanishes for  $t>0$.
In the studied problem, 
the instantaneous change of the Hamiltonian 
does not pump an  energy into the system.

Second,  employing (\ref{alfaG}), we find  
\be
 \lara{\psi(t)|H_0|\psi(t)}= -2\tilde{\cal E} - 2
 \sum_\KS\OM{k}|\lambda_\KS|^2 \cos(\OM{k}t) .
 \label{dcwihu}
\ee
Using the Riemann-Lebesgue lemma, 
we arrive at the following  general result 
\be
\lim_{t\to\infty} \lara{\psi(t)|H_0|\psi(t)} = -2\tilde{\cal E}
\label{Ex00}
\ee
 holding for finite  $\tilde{\cal E}$; see Appendix~\ref{RL}. 
 Due to (\ref{H0H1}),  it implies
\be
\label{Ex11}
\lim_{t\to\infty}   \lara{\psi(t)|H_1(0;t)|\psi(t)}=2\tilde{\cal E}.
\ee
Note that there is a  fundamental difference
between expanding (\ref{Ex00}) and static  (\ref{h0h1a})
despite the fact that 
expanding (\ref{Ex11}) and static  (\ref{h0h1b})
agree. This will  be explained below (\ref{dmkl}).

\subsection{Energy fluctuations}
\label{Fluctuations_sub}

The expectation values of powers of  the Hamiltonian
can be efficiently computed 
in the following way. 
We  consider $t>0$ and introduce 
\be
{\cal F}(s)=\lara{\psi(t)|\exp(-\ii H s )|\psi(t)},
\label{kdwew23}
\ee
where time-independent $H$ is given by (\ref{Hdiag}) and 
Schr\"odinger-picture  $|\psi(t)\ra$ is given by (\ref{SchSch}).
Then, we observe that $\exp(-\ii H s)$ is the 
Schr\"odinger-picture  evolution operator in the 
discussed setting, and so 
\be
{\cal F}(s)=\lara{\psi(t)|\psi(t+s)}.
\label{ewf2}
\ee
Using (\ref{SchSch}),  we  obtain 
\begin{align}
&{\cal F}(s)=\prod_\KS\exp(\Delta_\KS) 
\lara{\alpha_\KS(t)\exp(-\ii\OM{k} t)|\alpha_\KS(t+s)\exp\B{-\ii\OM{k} (t+s)}},\\
&\Delta_\KS= \ii \im[\Phi_\KS(t+s)-\Phi_\KS(t)],
\end{align}
where 
\begin{align}
&\lara{\alpha_\KS(t)\exp(-\ii\OM{k} t) |\alpha_\KS(t+s)\exp\B{-\ii\OM{k} (t+s)} } =\exp(\delta_\KS),\\
&\delta_\KS= -|\alpha_\KS(t)|^2/2 -   |\alpha_\KS(t+s)|^2/2
+\alpha^*_\KS(t)\alpha_\KS(t+s)\exp(-\ii\OM{k} s)
\end{align}
follows from (\ref{alfabeta}).
Employing  (\ref{alfaG}) and (\ref{PhiG}), we arrive 
at\footnote{As a self-consistency check, one may easily verify that 
$|{\cal F}(s)|\le1$ follows  from (\ref{wejf}).}
\be
{\cal F}(s)=\exp\B{\sum_\KS (\exp(-\ii s\OM{k})-1+\ii
s\OM{k})|\lambda_\KS|^2}.
\label{wejf}
\ee
Next, we consider  $\lara{H^n}\equiv\lara{\psi(t)|H^n|\psi(t)}$ 
and observe that when the evolution of $|\psi(t)\ra$
proceeds under (\ref{ewfkmlk}), then $H=H(0;0^+)=H(0;t)$ and so 
$\lara{H^n}=\lara{\psi(t)|H^n(0;t)|\psi(t)}$ quantifies energy 
fluctuations in the  state $|\psi(t)\ra$. On account 
of time independence of the Hamiltonian for $t>0$, $d\lara{H^n}/dt=0$ 
for $t>0$.
The idea now is to obtain $\lara{H^n}$ via 
\be
\lara{H^n}=\left.\ii^n \frac{d^n}{ds^n}{\cal F}(s)\right|_{s=0}.
\label{wslke}
\ee

Combining  (\ref{wejf}) and (\ref{wslke}), we arrive at 
\be
\lara{H}=0, \ \lara{H^2}=h_2, \ \lara{H^3}=h_3, \ \lara{H^4}=h_4+3 h_2^2,
\label{fhdfkl}
\ee
where 
\be
h_n = \sum_\KS \OM{k}^n |\lambda_\KS|^2.
\label{hnn}
\ee
From these expressions, we gather that  
the energy ${\cal E}=0$, in agreement with  (\ref{Hall}), while 
the standard deviation    of the energy is given by 
\be
\sigma_{\cal E}=\sqrt{\lara{H^2}-\lara{H}^2}=\sqrt{h_2}.
\label{sigE}
\ee
Moreover, the skewness and excess kurtosis read
\be
S_{\cal E}=\frac{\lara{(H-\lara{H})^3}}{\lara{(H-\lara{H})^2}^{3/2}}=\frac{h_3}{h_2^{3/2}},
\
K_{\cal E}=\frac{\lara{(H-\lara{H})^4}}{\lara{(H-\lara{H})^2}^2}-3=\frac{h_4}{h_2^2},
\label{sskk}
\ee
respectively.
Several remarks are in order now.

First,  (\ref{wejf}) can be also written as
\be
{\cal F}(s)=\exp\BB{\sum_\KS |\lambda_\KS|^2 \B{\sum_{n=2}^\infty \frac{(-\ii s\OM{k})^n}{n!}}},
\ee
and then  interchanging the order of the summations we  obtain
\be
\label{hnn0}
{\cal F}(s)\doteq \exp\B{\sum_{n=2}^\infty \frac{(-\ii s)^n}{n!} h_n},
\ee
where the dot over the equality symbol indicates that the expression
should be understood in a  formal sense because we shall not
analyze the conditions for interchanging the order of summations. 
Instead, we observe that the results listed in  (\ref{fhdfkl}) are 
obtained  irrespective of
whether (\ref{wejf}) or (\ref{hnn0}) is substituted into (\ref{wslke}).

Second, computing  $\lara{0|H^n|0}$,  we have verified
that the results listed in (\ref{fhdfkl}) are 
correct.\footnote{We have used the fact that $\lara{H^n}$ can be obtained from 
$\lara{\psi(0^+)|H^n(0;0^+)|\psi(0^+)}$, where $|\psi(0^+)\ra=|\psi(0)\ra=|0\ra$ and 
$H(0;0^+)=H$ is given by (\ref{Hdiag}).}
At the risk of stating the obvious, we mention that this method of computing $\lara{H^n}$  
is less efficient than the one based on (\ref{wslke}).

Third, we have some  general remarks about the above quantities. 
From a statistical viewpoint \cite{RoeBook}, 
 (\ref{kdwew23}) represents the
characteristic function,  $\lara{H^n}$ is
the $n$-th moment of the Hamiltonian, which 
coincides with the  $n$-th central moment due to $\lara{H}=0$,
and $h_n$ represents  the associated $n$-th cumulant
for $n\ge2$ (the first cumulant  vanishes).
In fact, for sufficiently small $n$, it can be verified using 
\cite{Mathematica143} that standard relations between
such defined  cumulants and moments are satisfied in the discussed setting.
Then, we observe that  (\ref{ewf2}) traditionally appears in the 
studies of quantum decay,
where it represents the survival amplitude \cite{DecayReview1978,Muga1996}. 
One of the goals there is to determine how the moments of the Hamiltonian
impact the dynamics of the survival amplitude. 
We explore in our studies 
the inverse approach, where the  knowledge of (\ref{ewf2}) is used to 
determine  the moments of the Hamiltonian.
Finally, we observe that (\ref{ewf2}) is discussed in the context 
of dynamical quantum phase transitions, where it is known as the 
Loschmidt amplitude \cite{Heyl2018}.

\subsection{Photon number}
\label{Photon_sec}

The number of photons in the state $|\psi(t)\ra$,
\be
{\cal N}(t)=\lara{\psi(t)|N|\psi(t)}
=\sum_\KS |\alpha_\KS(t)|^2,
\ee
reads 
\be
{\cal N}(t)=  2\tilde{\cal N} -2\sum_\KS |\lambda_\KS|^2\cos(\OM{k}t)
\label{Phhhhhh}
\ee
when (\ref{alfaG}) holds.
The standard deviation  of this   quantity, $\sigma_{\cal N}(t)$, 
defined analogously to   (\ref{VarN}), is 
equal to $\sqrt{{\cal N}(t)}$.
In the limit of $t\to\infty$, 
we proceed just as with  (\ref{dcwihu}),
 finding the following general result 
\be
{\cal N}(t\to\infty)=2\tilde{\cal N}
\label{Ntin}
\ee
 valid for finite $\tilde{\cal N}$; see Appendix~\ref{RL}.
The factor of two difference between $\tilde{\cal N}$ and ${\cal N}(t\to\infty)$
is reminiscent of the relation 
between (\ref{h0h1a}) and (\ref{Ex00}). However, one must keep in mind that
the energy represented  by the operator $H_0$ is a different object than the photon 
number represented  by the operator $N$.

\section{Sum-free expressions}
\label{Sum_sec}
Several  expressions, which  have already been introduced,  involve
the summation over momenta $\BD{k}$ and polarizations $\sigma$.
The former can be  replaced by the  integral
\begin{subequations}
\be
\sum_{\BD{k}}\frac{1}{V} \mapsto \int \frac{d^3k}{(2\pi)^3},
\label{sumINT}
\ee
whereas the latter  can be carried out via 
the completeness relation \cite{Greiner}
\be
\sum_\sigma \eta^i(\BD{k},\sigma)
\eta^j(\BD{k},\sigma)=\delta_{ij}-\frac{k^i k^j}{\OM{k}^2}.
\label{dwefe}
\ee
The resulting expressions can be simplified  using
\be
\BD{k}\cdot\BD{J}_\text{mat}(\BD{k})=0,
\label{coNS}
\ee
\label{SF}%
\end{subequations}
which follows  from   (\ref{nabjmat}) and removes
the contribution of the second term  
of (\ref{dwefe}) to the discussed quantities.

Using  (\ref{SF}), we have found that 
(\ref{VVVstat}), (\ref{tildeE}),  and (\ref{Nph})
can be also written as   
(\ref{st1}), (\ref{st2}), and (\ref{st3}), respectively.
By the same token, (\ref{VVV}),  (\ref{dcwihu}),  (\ref{hnn}), and (\ref{Phhhhhh})
 are equivalent to 
(\ref{ne1}), (\ref{ne2}), (\ref{ne3}), and (\ref{ne4}),
respectively.

\section{Static magnetic dipole field}
\label{Static_sec2}
To illustrate the results of Sec. \ref{Static_sec},  
we consider the following   external current 
satisfying (\ref{nabjmat})
\be
\jmat(\BD{r})=  \BD{\mu}\times\BD{r} f(\BD{r}),
\label{JMATMAG}
\ee
 where 
\be
f(\BD{r}) = \frac{2}{\epsilon^2}\delta_\epsilon(\BD{r}),
\ \delta_\epsilon(\BD{r})=
\frac{\exp\B{-r^2/\epsilon^2}}{\B{\sqrt{\pi}\epsilon}^3}
\ee
with $r=|\BD{r}|$. The three-dimensional nascent delta function $\delta_\epsilon(\BD{r})$
confines the current to a  volume of order  $\epsilon^3$ around $\BD{r}=\0$.
The prefactor  $2/\epsilon^2$ is chosen such that $\BD{\mu}$
represents the magnetic moment according to the standard  
formula. Namely,
$\int d^3r\, \BD{r}\times \BD{j}_\text{mat}(\BD{r})/2$, which can be e.g.
found in \cite{JacksonBook}, yields 
 $\BD{\mu}$ in the discussed problem.
We note that (\ref{JMATMAG}) 
is  qualitatively the same as in a  current-in-a-loop-wire setup.
We also note that we need for practical calculations
\be
\label{JmatS}
\BD{J}_\text{mat}(\BD{k}) =-\ii \BD{\mu}\times\BD{k}\exp\B{-\frac{\epsilon^2\OM{k}^2}{4}}.
\ee
This expression is used in Appendix \ref{STAT_app} to evaluate the integrals 
determining (\ref{tildeEE}), (\ref{AvecS}), and (\ref{tildeNN}).
Moreover,  (\ref{JmatS}) shows that the parameter $1/\epsilon$ sets the
effective ultraviolet (UV) cutoff in the problem under consideration.
Indeed, the Gaussian factor in (\ref{JmatS}),
which enters the integrals determining (\ref{tildeEE}), (\ref{AvecS}), and (\ref{tildeNN}), 
rapidly approaches zero  for photon momenta whose magnitude exceeds
$1/\epsilon$.\footnote{Several quantities discussed in this work are divergent
for $\epsilon\to0^+$.
This UV divergence is an expected feature because it
arises as the effective support of the external current shrinks to a point.}

\begin{figure}[t]
\includegraphics[width=\pref\columnwidth,clip=true]{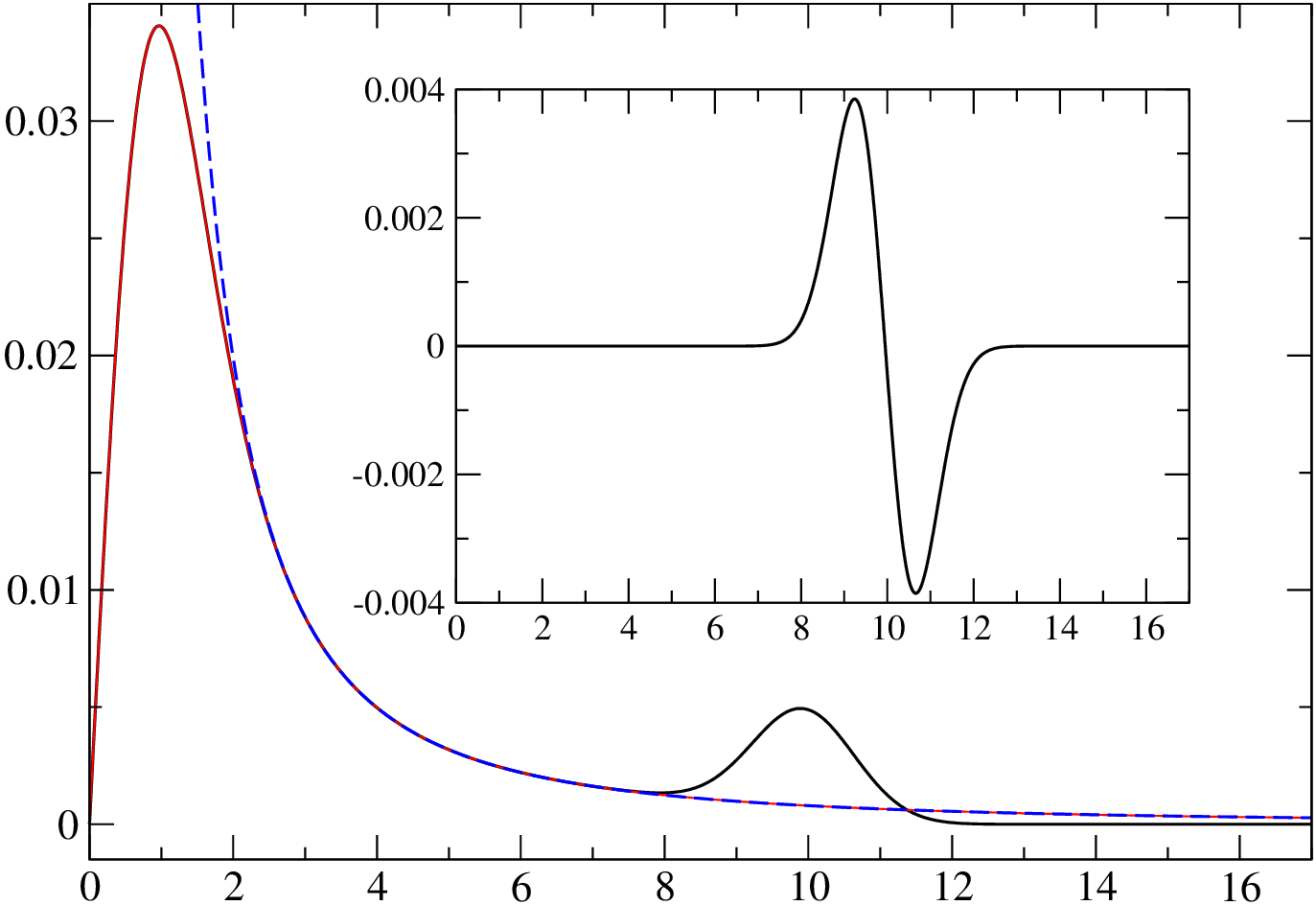}
\caption{The main plot shows 
$\epsilon^{2} \times \lara{\BD{A}}\cdot ( \BD{\mu}\times\rhat)/| \BD{\mu}\times\rhat|^2$
as a function of $r/\epsilon$.
The solid black line represents  expanding result (\ref{meanA})  for $t/\epsilon=10$, 
the solid red line shows 
static result (\ref{AvecS}), 
and the dashed blue line corresponds to    point magnetic dipole
result (\ref{jlfw}).
The solid  black line quickly approaches the null result for 
$r/\epsilon>t/\epsilon+2$.
The solid black and red (solid red and dashed blue)  lines  overlap well 
for $r/\epsilon< t/\epsilon-2$
($r/\epsilon> 2$).
The inset shows 
$\epsilon^{3}\times \lara{\BD{E}}\cdot ( \BD{\mu}\times\rhat)/|\BD{\mu}\times\rhat|^2$
as a function of $r/\epsilon$. It is based on (\ref{EEexp}) for  $t/\epsilon=10$;
the null result is quickly approached for $|r/\epsilon-t/\epsilon|>2$.
The results presented in the main plot and the inset are qualitatively the same 
for  longer  times ($t/\epsilon>10$).
}
\label{mag_fig}
\end{figure}

To begin the discussion, we observe that  the energy of the (ground) state 
$|\tilde{\psi}(t)\ra$ equals
\be
\tilde{\cal E}=-\frac{\mu^2}{3(\sqrt{2\pi}\epsilon)^3},
\label{tildeEE}
\ee
where $\mu=|\BD{\mu}|$.
The vector field in this state reads
\be
\lara{\tilde{\psi}(t)|\BD{A}(0,\BD{r})|\tilde{\psi}(t)}=
\frac{\BD{\mu}\times\rhat}{4\pi r^2}
\erf\B{\frac{r}{\epsilon}} -\frac{ \BD{\mu}\times\rhat  }{2\pi r}\delta_\epsilon(r),
\label{AvecS}
\ee
where
\begin{align}
\label{erfX}
&\erf(x)=\frac{2}{\sqrt{\pi}}\int_0^x dy \exp(-y^2),\\
&\delta_\epsilon(x)=\frac{\exp\B{-x^2/\epsilon^2}}{\sqrt{\pi}\epsilon}
\end{align}
are the error function and
the  one-dimensional nascent delta function, respectively.
For $r\gg\epsilon$, the second  term in (\ref{AvecS}) is
negligible while the first one can be written as 
\be
 \frac{\BD{\mu}\times\rhat}{4\pi r^2},
\label{jlfw}
\ee
which is the standard  result for the vector field of a  magnetic dipole 
in the long-distance limit.
The second term in (\ref{AvecS}) is non-negligible for $r=O(\epsilon)$.
In particular, its contribution is essential to ensure that
the vector field remains finite at  $\BD{r}=\0$.
Equations (\ref{AvecS}) and (\ref{jlfw}) are shown and 
compared in Fig. \ref{mag_fig}.

The magnetic field in the state $|\tilde{\psi}(t)\ra$, 
given by the curl of (\ref{AvecS}),  reads 
\be
\lara{\tilde{\psi}(t)|\BD{B}(0,\BD{r})|\tilde{\psi}(t)}=
\frac{3(\BD{\mu}\cdot\rhat)\rhat-\BD{\mu}}{4\pi r^3}
\erf\B{\frac{r}{\epsilon}}
+\BD{\mu}\delta_\epsilon(\BD{r})
-\frac{3(\BD{\mu}\cdot\rhat)\rhat-\BD{\mu}}{2\pi r^2}\delta_\epsilon(r)
+\frac{(\BD{\mu}\cdot\rhat)\rhat}{2\pi r}\partial_r\delta_\epsilon(r).
\label{adcnm}
\ee
The first term in (\ref{adcnm}) reproduces the standard
result for $r\gg \epsilon$. The remaining terms in (\ref{adcnm}),   
when integrated over all space, yield $2\BD{\mu}/3$, which is consistent with the
well-known $2\BD{\mu}\delta(\BD{r})/3$ 
contribution to the magnetic field of a point magnetic dipole 
(see e.g. \cite{JacksonBook}).

Finally, we find it interesting that the number of photons building
the discussed  state is finite
\be
\tilde{\cal N}= \frac{1}{6\pi^2} \B{\frac{\mu}{\epsilon}}^2,
\label{tildeNN}
\ee
there is no IR catastrophe so to speak.
It should be noted that the  finiteness of 
$\tilde{\cal E}$ does not 
guarantee the finiteness of $\tilde{\cal N}$
($\tilde{\cal E}<\infty$ does not exclude the possibility 
of the IR divergence of $\tilde{\cal N}$).

\section{Expanding  magnetic dipole field}
\label{Expanding_sec}

To illustrate the results of Sec.~\ref{Problem_sec}, we assume now that 
$\jmat(\BD{r})$ is given by (\ref{JMATMAG}), and so we study  
the dynamics induced by the following   time-dependent 
 current satisfying (\ref{nabjmat}) 
\be
\jmat(t,\BD{r})=  \BD{\mu}\times\BD{r} f(\BD{r})\theta(t).
\label{JmatMAGt}
\ee
We discuss below several quantities characterizing the
state $|\psi(t)\rangle$ representing the expanding magnetic dipole field.
The integrals entering (\ref{meanA}), (\ref{dmkl}), (\ref{wdfd3}),
and (\ref{Ntdaw}) are evaluated in Appendix \ref{NONEQ_app}.
Moreover, till the end of this section, we assume 
\be
t\ge0.
\ee

\subsection{Vector, magnetic, and electric fields} 
We have found 
\begin{subequations}
\begin{align}
\lara{\psi(t)|\BD{A}(0,\BD{r})|\psi(t)} &= \frac{\BD{\mu}\times\rhat}{4\pi r^2}
\BB{\erf\B{\frac{r}{\epsilon}}
-\frac{1}{2}\erf\B{\frac{r-t}{\epsilon}}
-\frac{1}{2}\erf\B{\frac{r+t}{\epsilon}}} \label{meanAa} \\
&-\frac{ \BD{\mu}\times\rhat  }{2\pi r}\delta_\epsilon(r) \label{meanAb}\\
&+\frac{ \BD{\mu}\times\rhat  }{4\pi r}\BB{\delta_\epsilon(r-t)+\delta_\epsilon(r+t)}
 \label{meanAc},
\end{align}
\label{meanA}%
\end{subequations}
which is shown in Fig. \ref{mag_fig}.
Equation (\ref{meanAa}) describes the  expanding vector field 
of the magnetic dipole. This is best seen in the $\epsilon\to0^+$ 
limit, where (\ref{meanAa})   can be written for $r>0$ 
as\footnote{Equations (\ref{aaAPPROX}) and  
(\ref{dewnkl}) rest upon the  assumption that $\theta(0)=1/2$.} 
\be
\frac{\BD{\mu}\times\rhat}{4\pi r^2}\theta(t-r)
\label{aaAPPROX}
\ee
with the help of
\be
\lim_{\epsilon\to0^+}\erf\B{\frac{x}{\epsilon}} 
=\theta(x)-\theta(-x).
\ee
Note that in this  limit there is a discontinuous 
shockwave-like  front 
at $r=t$ spreading  information about the source of the 
magnetic dipole field.
As this process causally proceeds with the speed of 
light, this front may be termed a causal front.
For $r\gg\epsilon$ but without assuming $\epsilon\to0^+$,
(\ref{aaAPPROX}) is valid for $|r-t|\gg\epsilon$
and 
there are some  corrections to 
(\ref{aaAPPROX}) in the neighborhood of $r=t$.
Namely, there is a crossover region near $r=t$, where the vector
field smoothly 
goes from $\BD{\mu}\times\rhat/(4\pi r^2)$ to $\0$.
This is the consequence of the fact that for a fixed 
$\epsilon>0$, the  magnetic dipole is created in an extended 
region of space at $t=0$
(its  field is then causally propagated).
Equation (\ref{meanAb}), which is localized around $\BD{r}=\0$ and 
represents a static contribution, 
is the same as in Sec. \ref{Static_sec2}. Finally, (\ref{meanAc})
is localized  near $r=t$ and represents a propagating 
contribution. Note that (\ref{meanAc}), just as (\ref{meanAa}),
encodes  a shockwave-like  structure,  which is clearly seen in the limit of
$\epsilon\to0^+$.

The  magnetic field created by (\ref{JmatMAGt}) reads
\begin{subequations}
\begin{align}
\lara{\psi(t)|\BD{B}(0,\BD{r})|\psi(t)} &=\frac{3(\BD{\mu}\cdot\rhat)\rhat-\BD{\mu}}{4\pi r^3}
\BB{\erf\B{\frac{r}{\epsilon}}
-\frac{1}{2}\erf\B{\frac{r-t}{\epsilon}}
-\frac{1}{2}\erf\B{\frac{r+t}{\epsilon}}
} \label{meanBa}\\
&+\BD{\mu}\delta_\epsilon(\BD{r})
-\frac{3(\BD{\mu}\cdot\rhat)\rhat-\BD{\mu}}{2\pi r^2}\delta_\epsilon(r)
+\frac{(\BD{\mu}\cdot\rhat)\rhat}{2\pi r}\partial_r\delta_\epsilon(r)
\label{meanBb}\\
&+\frac{\BD{\mu}}{4\pi} \partial_r\B{\frac{\delta_\epsilon(r-t)+\delta_\epsilon(r+t)}{ r}}
-\frac{(\BD{\mu}\cdot\rhat)\rhat}{4\pi} r^2
 \partial_r\B{\frac{\delta_\epsilon(r-t)+\delta_\epsilon(r+t)}{ r^3}}
 \label{meanBc}.
\end{align}
\label{meanB}%
\end{subequations}
As this  field is given by the curl  of (\ref{meanA}),
(\ref{meanB}) can be discussed similarly to (\ref{meanA}). Thereby, 
we shall not dwell on it. We only mention that 
 (\ref{meanBa}) for $r>0$ and $\epsilon\to0^+$ reads
\be
\frac{3(\BD{\mu}\cdot\rhat)\rhat-\BD{\mu}}{4\pi r^3}\theta(t-r),
\label{dewnkl}
\ee
(\ref{meanBb}) is the same as in Sec. \ref{Static_sec2}, 
and (\ref{meanBc}) is qualitatively similar to  (\ref{meanAc}).

The electric field created by (\ref{JmatMAGt}), obtained by acting with
$-\partial_t$   on (\ref{meanA}), reads
\be
\lara{\psi(t)|\BD{E}(0,\BD{r})|\psi(t)} =\frac{\BD{\mu}\times\rhat}{4\pi} \partial_r
\B{\frac{\delta_\epsilon(r-t)-\delta_\epsilon(r+t)}{ r}},
\label{EEexp}
\ee
which is shown in Fig. \ref{mag_fig}.
As the electric field vanishes for a static   magnetic dipole, (\ref{EEexp}) should 
vanish for $t-r\gg\epsilon$. Moreover, (\ref{EEexp})
should also vanish for $r-t\gg\epsilon$ for causal reasons. Thereby, it can be 
non-zero only near $r=t$, which we see from the above expression. In fact, it
must be non-zero near $r=t$  because the magnetic field 
is time dependent there. A shockwave-like structure is exhibited 
by (\ref{EEexp}), which is  clearly seen in the limit of
$\epsilon\to0^+$.

Finally, we note that the expanding solution discussed in this
work bears some similarity to the incoming (outgoing) 
solutions considered in our  studies of the Proca theory, 
where  vector bosons are massive (see e.g. \cite{BDDipole1}).

\subsection{Overlap} 
Combining (\ref{overLAP}) with (\ref{tildeNN}), we have obtained
\be
\lara{\tilde{\psi}(t)|\psi(t)} = \exp\B{ -\frac{\mu^2}{12(\pi\epsilon)^2}}. 
\label{Ovv}
\ee
Two  remarks are in order now.

First, as time increases, 
the expanding magnetic dipole field increasingly resembles 
the static magnetic dipole field.
From this perspective, time independence of the overlap between the states 
representing these fields is counterintuitive.
Second, given that the long-distance features of the expanding and static
magnetic dipole fields fundamentally differ, one might expect
an IR singularity in the argument of the exponent in
(\ref{Ovv}), leading to the vanishing overlap. From this viewpoint,
the fact that (\ref{Ovv}) is  non-zero is
counterintuitive.

\subsection{Energy}
We have found 
\be
\lara{\psi(t)|H_0|\psi(t)}=-2\tilde{\cal E}
\B{1-\exp\B{-\frac{t^2}{2\epsilon^2}}
+\B{\frac{t}{\epsilon}}^2\exp\B{-\frac{t^2}{2\epsilon^2}} },
\label{dmkl}
\ee
where $\tilde{\cal E}$ is given by (\ref{tildeEE}).
This result determines $\lara{\psi(t)|H_1(0;t)|\psi(t)}$
via (\ref{H0H1}) and confirms general prediction (\ref{Ex00}).
We are ready now to explain the remark 
below (\ref{Ex11}).
For $t\to\infty$, the expectation value of the 
interaction  Hamiltonian  in the states representing  static and expanding 
fields  is the same because 
the vector field of the expanding solution 
takes a static value in the volume, where the external 
current is localized. Thereby, due to (\ref{Hall}),
the {\it excess} energy with respect to the static solution, 
$-\tilde{\cal E}=|\tilde{\cal E}|$,
must be deposited in the expectation value of the free-field 
Hamiltonian for $t\to\infty$. 
As far as the spatial localization of this  
energy is concerned, we observe that it must be 
localized in the neighborhood of $r=t$.
We have verified this  observation 
numerically integrating  the  energy density.

Finally, we note the dual role of the shockwave-like front in 
the causal propagation  of the magnetic
dipole field. On the
one hand, it   spreads 
information  about  the 
source of the magnetic dipole 
field. On the other hand, it
carries the amount of energy required 
to keep the total energy null (\ref{Hall}).

\subsection{Energy fluctuations}
We have obtained
\be
h_n=\frac{\mu^2}{6\pi^2} 
\frac{2^{n/2}\Gamma(1+n/2)}{\epsilon^{n+2}},
\label{wdfd3}
\ee
where $\Gamma$ is the gamma function.
Using  (\ref{fhdfkl}), this result  leads to 
\be
\lara{H}=0, \ \lara{H^2}=\frac{\mu^2}{3\pi^2\epsilon^4}, 
\ \lara{H^3}=\frac{\mu^2}{(2\pi)^{3/2}\epsilon^5},\
\lara{H^4}=\frac{4\pi^2\epsilon^2\mu^2+\mu^4}{3\pi^4\epsilon^8}.
\label{jjh}
\ee
Thereby, the standard deviation   of the energy is
\be
\sigma_{\cal E}=\frac{\mu}{\sqrt{3}\pi\epsilon^2}.
\label{EsigE}
\ee
We stress that there are  no energy fluctuations in the state $|\tilde{\psi}(t)\ra$,
representing the static  magnetic dipole field in the setup discussed 
in Sec. \ref{Static_sec2}; $\tilde{\sigma}_{\cal E}=0$.
Finally,  for the sake of completeness, we also provide expressions 
for the skewness
and excess kurtosis 
\be
S_{\cal E}=\B{\frac{3}{2}\pi}^{3/2}\frac{\epsilon}{\mu},
\ K_{\cal E}=12\pi^2\B{\frac{\epsilon}{\mu}}^2,
\ee
which follow from (\ref{jjh}) via  (\ref{sskk}).

\subsection{Photon number}
We have found 
\be
\label{Ntdaw}
{\cal N}(t)= \frac{1}{3\pi^2}\B{\frac{\mu}{\epsilon}}^2
2\frac{t}{\sqrt{2}\epsilon}\daw\B{\frac{t}{\sqrt{2}\epsilon}},
\ee
where
\be
\daw(x)=\int_0^x dy \exp\B{y^2-x^2},
\label{dawx}
\ee
is the Dawson  function.
Two  remarks are in order now.

\begin{figure}[t]
\includegraphics[width=\pref\columnwidth,clip=true]{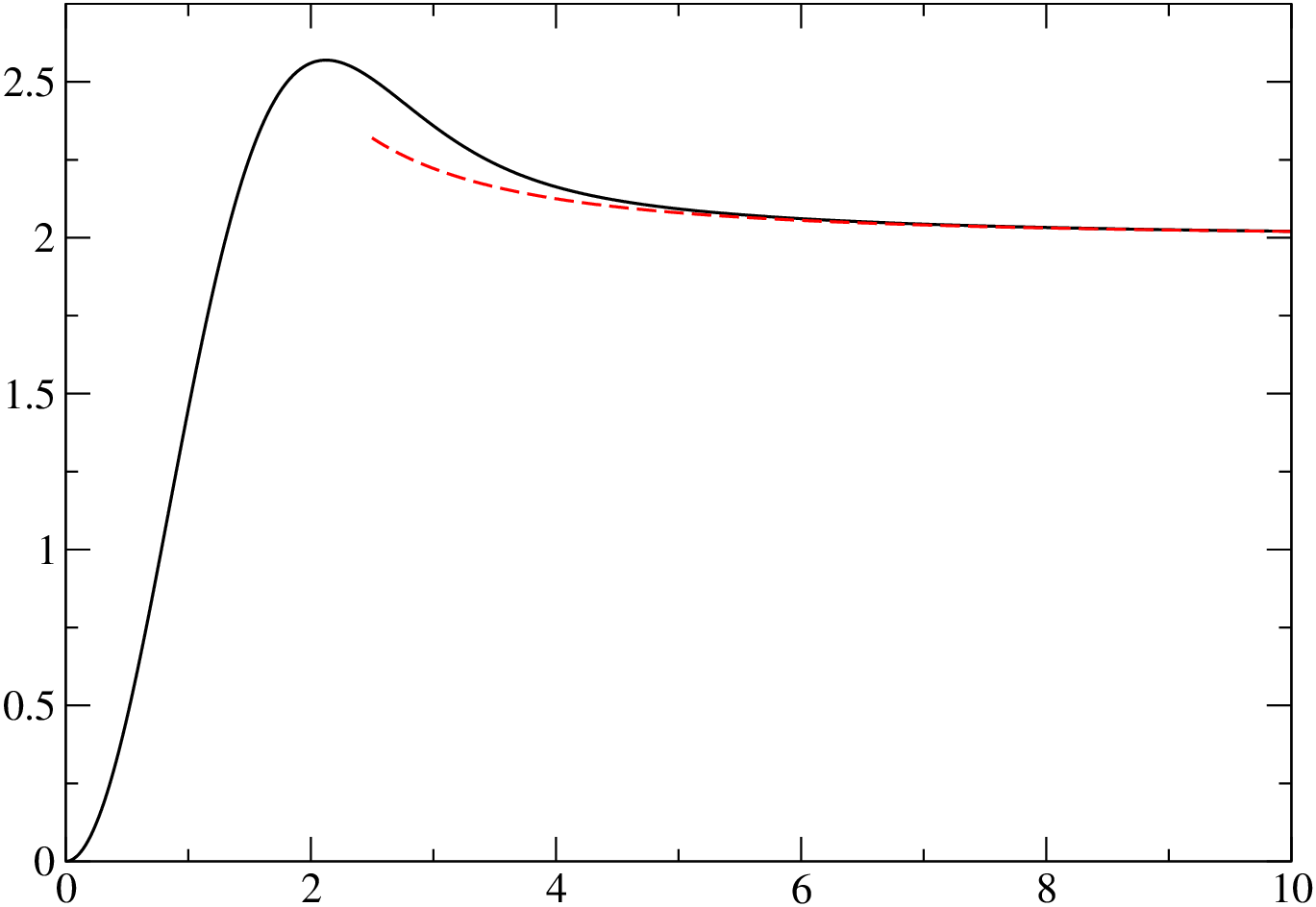}
\caption{The rescaled photon number, ${\cal N}/\tilde{\cal N}$, 
as a function of $t/\epsilon$. The solid black line represents 
(\ref{Ntdaw}) while the  red dashed line shows (\ref{Nalg}).
}
\label{NAss}
\end{figure}

First, given that 
\be
\daw(x)\simeq x \for x\ll 1,
\label{daw1}
\ee
${\cal N}(t)$ vanishes for $t=0$, which is  expected.
More interestingly,
\be
\daw(x)\simeq \frac{1}{2x} + \frac{1}{4x^3} \for x\gg1
\label{daw2}
\ee
leads to
\be
{\cal N}(t)\simeq 2\tilde{\cal N}
+2\tilde{\cal N}\B{\frac{\epsilon}{t}}^2 \for t\gg\epsilon,
\label{Nalg}
\ee
which extends  prediction (\ref{Ntin}) to finite times. The derivations of 
(\ref{daw1}) and (\ref{daw2}) are briefly discussed in 
Appendix~\ref{Dawson}.  The relation  between $\tilde{\cal N}$ and
${\cal N}(t\to\infty)$ suggests that $\tilde{\cal N}$ photons build the
shockwave-like front, again no IR catastrophe is found in the photon number.

Second, ${\cal N}(t)$   is depicted in Fig.
\ref{NAss}.
We see there that photons are radiated by the external 
current.
Instead of a sudden photon burst at the moment of magnetic dipole creation,
the number of radiated photons gradually increases at early times.
At $t\simeq 2.1 \epsilon$, the maximum 
in the number of photons is reached, and then a  steady state is
approached  for $t\gg\epsilon$. This does not mean that 
the occupation of individual $\KS$ modes ceases to evolve for $t\to\infty$.
Indeed, this quantity is proportional to 
$|1 - \exp(\ii\OM{k}t)|^2$, implying continuous emission 
and absorption of photons by the external source.

\section{Orders of magnitude}
\label{Orders_sec}

The question now is what are the orders of magnitude of the discussed
quantities. To address this issue, we 
assume  that  $\mu=\beta\mu_B$, 
where $\mu_B$ is the Bohr's magneton. Then, dimensionless (\ref{tildeNN})
and (\ref{Ovv}) can be conveniently computed  via
\be
\frac{\mu}{\epsilon} =
\beta \sqrt{\frac{\alpha}{4\pi}}\frac{\lambda}{\epsilon},
\label{mueps2}
\ee
where $\alpha\simeq1/137$ is the fine-structure constant while 
$\lambda\simeq2.4\times10^{-12}\,\text{m}$ is the Compton wavelength of the
electron.
By proceeding similarly, (\ref{tildeEE}) and (\ref{EsigE}) can also be computed.
We now examine  two vastly different examples.

\begin{table}[t]
\centering
\begin{tabular}{c|c}
 expanding & static \\
\hline
$ {\cal E}=0 $ & $\tilde{\cal E}\simeq-2.8\times10^{-2}\, m_e c^2$  \\
\hline
$\sigma_{\cal E}\simeq 4.2\, m_e c^2$ & $\tilde{\sigma}_{\cal E}=0$
\\
\hline
${\cal N}(t\to\infty)\simeq1.2\times10^{-4}$ 
& $\tilde{\cal N}\simeq5.8\times10^{-5}$
\\
\hline
$\sigma_{\cal N}(t\to\infty)\simeq1.1\times10^{-2}$ & 
$\tilde{\sigma}_{\cal N}\simeq7.6\times10^{-3}$
\end{tabular}
\caption{Comparison of the expanding and static cases for the microscopic magnet.}
\label{NeutTab}
\end{table}

First, we consider the {\it microscopic}  magnet choosing $\epsilon=1\,\text{fm}$ and $\beta=10^{-3}$ 
to estimate the quantities of interest in the regime of parameters 
relevant for a neutron. The results are given in Table \ref{NeutTab},
where $m_e$ is the electron mass and
the factor of $c^2$ is inserted so that 
$\tilde{\cal E}$ and    $\sigma_{\cal E}$ are expressed in the 
conventional form.
The overlap between the states representing 
static and expanding fields is approximately $0.99997$.

Second, we consider the {\it macroscopic} 
magnet choosing  $\epsilon=0.01\,\text{m}$ and $\beta=7.7\times10^{23}$.
This  value of $\beta$ arises  when one assumes that the magnetic moment 
$\mu$ is the same as that of the maximally-magnetized
iron ball of radius $0.01\,\text{m}$ at  room 
temperature.\footnote{A density of $7874\,\text{kg}\,\text{m}^{-3}$ and a
saturation magnetization of $217.6\,\text{J}\text{T}^{-1}\text{kg}^{-1}$ are
assumed for the iron ball (see \cite{magFe} for the latter).}
The results are  given 
in Table \ref{IronTab}, where joule, instead of $m_e c^2$, is 
used to report  $\tilde{\cal E}$ and    $\sigma_{\cal E}$ 
because we discuss a  macroscopic   object now.
The overlap between the states representing static and expanding fields 
does not vanish but is extraordinarily close to zero 
(its natural logarithm is of the order of $-10^{23}$).

The above calculations  show that  energy- and photon number-related 
quantities appearing in our calculations 
are neither astronomically large nor microscopically small.

\begin{table}[t]
\centering
\begin{tabular}{c|c}
 expanding & static \\
\hline
$ {\cal E}=0$ & $ \tilde{\cal E}\simeq-1.4\,\text{J}$    \\
\hline
$\sigma_{\cal E}\simeq
2.6\times10^{-12}\,\text{J} $ & $\tilde{\sigma}_{\cal E}=0$
\\
\hline
${\cal N}(t\to\infty)\simeq6.9\times10^{23}$ 
& $\tilde{\cal N}\simeq3.4\times10^{23}$
\\
\hline
$\sigma_{\cal N}(t\to\infty)\simeq8.3\times10^{11}$ & 
$\tilde{\sigma}_{\cal N}\simeq5.9\times10^{11}$
\end{tabular}
\caption{Comparison of the expanding and static cases for the macroscopic magnet.}
\label{IronTab}
\end{table}

\section{Summary}
\label{Summary_sec}

The expanding quantum magnetic field, 
whose theory is developed in this work and whose properties are briefly summarized  below,
appears when an external current is instantaneously switched on and subsequently remains constant in time.
This  field features the 
shockwave-like front that propagates at the speed of light.
Within the region it has swept through, a quantum magnetostatic field is locally 
established,
whereas  in the region it has not yet reached, 
no field associated with the discussed field  source is present.
We have contrasted this behavior with that of the corresponding 
quantum magnetostatic field, which extends throughout all space
and exhibits no expanding characteristics.
In particular, by studying  energy- and photon number-related observables as
well as their fluctuations, 
we have shown that  quantum states representing 
the expanding field and its magnetostatic
counterpart fundamentally differ.

As far as the energy is concerned, we have found that the
total energy of the expanding  field vanishes. 
This  quantity has two components. The 
one associated with  the electromagnetic field
and the one taking into account the  interaction of a
source with the field that it creates.
At first sight,  the vanishing total energy might suggest that
the expanding field can form spontaneously if no conservation 
laws are violated in the process. 
However, this interpretation overlooks
the fact that the energy required to create the field source 
has not been taken into account.
Interestingly, in the case of a magnetostatic 
field, the  total energy is negative but causality excludes
spontaneous formation of this field irrespective of the value of the
energy required to create the field source.
As the above remarks indicate, the energy of the expanding
field is higher than the energy of the corresponding  magnetostatic field.
This intuitively plausible feature reflects the energetic cost 
associated with the creation of the shockwave-like  front.
The photon number considerations offer an additional insight into 
this process. 
Specifically, the number of photons in the expanding  field 
is twice larger than in the corresponding magnetostatic field, 
which can be attributed to the presence of 
the shockwave-like  front.

The question now is what are physical consequences of 
the existence of the  shockwave-like  front.
First, 
it modifies the motion  of charged particles via 
the Lorentz force. In this respect, we would like to 
underscore the fact that besides a  magnetic field, there
is also an electric field in the shockwave-like  front.
Both  fields are inhomogeneous and encode 
details of a field source.
Second, the  shockwave-like  front carries the energy, which is 
equal to the modulus of the energy of the corresponding  
magnetostatic field. This energy can be transferred to systems 
interacting with it.
Finally, it is of physical interest to note  that for a magnetic field source
appearing at some instant and then being static, magnetostatic results work very well 
sufficiently close to the source but fail fundamentally in the asymptotic region.
In this  sense, magnetostatic results can be seen as 
a convenient fiction.

\acknowledgments

I am indebted to  Adolfo del Campo for   comments  about this work. 
The research for this publication has been supported by Jagiellonian University.

\appendix

\section{Conventions}
\label{Conventions_app}

The Heaviside-Lorentz system of units is used
in its $\hbar=c=1$ version throughout
(except in Sec.~\ref{Orders_sec},
where SI units  are employed when  needed). 
Vectors are written in bold.
The complex (hermitian) conjugation is denoted by $*$ and $\cc$ 
($\dag$ and $\hc$). The symbol $\equiv$ denotes a definition, 
$:\,:$ designates the  normal ordering,
$\partial_X=\partial/\partial X$,
 $x^+$  stands for  a quantity infinitesimally
larger  than $x$,  and $\BD{\hat{x}}=\BD{x}/|\BD{x}|$.

\section{Expectation values of free-field and interaction Hamiltonians in  excited eigenstates of static
problem}
\label{App_Virial}

The excited eigenstates of (\ref{Hdiag}) can be written as
\be
|\{n(\KS)\}\ra= \prod_\KS \frac{(\tilde{b}_\KS^\dag)^{n(\KS)}}{\sqrt{n(\KS)!}}|\tilde{0}\ra,
\label{nijw}
\ee
where $n(\KS)=0,1,2,\dots$ and at least one such quantity is non-zero.
The distribution of the energy between free-field and interaction Hamiltonians 
in these states reads 
\begin{align}
\label{dkpoe}
&\lara{\{n(\KS)\}|H_0|\{n(\KS)\}}=-\tilde{\cal E}+\sum_\KS n(\KS) \OM{k},\\
&\lara{\{n(\KS)\}|H_1|\{n(\KS)\}}=2\tilde{\cal E}.
\label{uygtvy}
\end{align}
Comparing (\ref{dkpoe}) and (\ref{uygtvy}) to (\ref{h0h1a}) and 
(\ref{h0h1b}), we see that the expectation value of $H_0$ ($H_1$) 
in the excited eigenstates disagrees (agrees) with the ground state value.

To derive the above results, we express $H_0=\sum_\KS \OM{k} b_\KS^\dag b_\KS$ in terms of
 $\tilde b_\KS$ and
$\tilde b_\KS^\dag$, which leads to    (\ref{dkpoe}). 
Then, we use (\ref{Hdiag}) to obtain  
\be
\lara{\{n(\KS)\}|H|\{n(\KS)\}}=\tilde{\cal E}+ \sum_\KS n(\KS) \OM{k}.
\ee
Since this quantity is the sum of the expectation values of free-field and interaction Hamiltonians,
and the former is given by (\ref{dkpoe}), it is evident  
that  (\ref{uygtvy}) holds.

\section{Long-time limit of (\ref{dcwihu}) and (\ref{Phhhhhh})}
\label{RL}
Expressions (\ref{dcwihu}) and (\ref{Phhhhhh}) can be written as 
${\cal I}(0)-{\cal I}(t)$, where 
\be
{\cal I}(t)=\sum_\KS\frac{f_\sigma(\BD{k})}{V}\cos(\OM{k}t)
\ee
with  $f_\sigma$ being a non-negative  function. 
Using (\ref{sumINT}),  this can be transformed into 
\be
{\cal I}(t)= \int_0^\infty  d\OM{k} \tilde{f}(\OM{k})\cos(\OM{k}t),
\ee
where the dependence of non-negative $\tilde{f}$ on $\OM{k}$  only
is guaranteed by an  angular integration in  momentum space.
Vanishing of 
${\cal I}(t\to\infty)$ follows from the Riemann-Lebesgue lemma 
when $\tilde{f}\in L^1(0,\infty)$, which for non-negative 
$\tilde{f}$  amounts to 
$\int_0^\infty  d\OM{k} \tilde{f}(\OM{k})={\cal I}(0)<\infty$.

\section{Integrals for  static problem}
\label{STAT_app}

{\bf The  useful integrals}. We will rely below on 
\begin{align}
\label{dOmega}
&\int
d\Omega(\khat)\exp(\ii\BD{k}\cdot\BD{r})=4\pi\frac{\sin(\OM{k}r)}{\OM{k}r},\\
&\int_0^\infty dx x^m\exp(-x^2/2)=2^{(m-1)/2} 
\Gamma\B{\frac{m+1}{2}},
\label{iiii}
\end{align}
where the former (latter) result follows (comes) from formula 
3.915.1 (3.461.2b) of \cite{Ryzhik}. At the risk of stating the obvious, 
we note that $\Omega(\khat)$ denotes  a solid angle in momentum space.

{\bf The vector field}. Combining  (\ref{lLKS}),   (\ref{VVVstat}),  and    (\ref{SF}), we arrive at 
\be
\label{st1}
\lara{\tilde{\psi}(t)|\BD{A}(0,\BD{r})|\tilde{\psi}(t)}=\int
\frac{d^3k}{(2\pi)^3}\frac{\BD{J}_\text{mat}(\BD{k})}{2\OM{k}^2}\exp(\ii\BD{k}\cdot\BD{r})+\cc
\ee
For (\ref{JmatS}), we find using  (\ref{dOmega}) that 
\begin{align}
&\lara{\tilde{\psi}(t)|\BD{A}(0,\BD{r})|\tilde{\psi}(t)}=
-(\BD{\mu}\times\bnabla)\B{\frac{I}{2\pi^2 r}},\\
&I=\int_0^\infty dx \frac{\exp(-x^2/4)}{x}\sin\B{x\frac{r}{\epsilon}},
\end{align}
where according to formula 3.952.6 of \cite{Ryzhik}
\be
I=\frac{\pi}{2}\erf\B{\frac{r}{\epsilon}}.
\ee
This leads to (\ref{AvecS}).

{\bf The energy}. Combining (\ref{lLKS}),  (\ref{tildeE}),  and (\ref{SF}), we arrive at
\be
\label{st2}
\tilde{\cal E}=-\int \frac{d^3k}{(2\pi)^3}\frac{\BD{J}_\text{mat}(\BD{k})\cdot\BD{J}^*_\text{mat}(\BD{k})}{2\OM{k}^2}.
\ee
For (\ref{JmatS}), we have 
\be
\tilde{\cal E}= -\frac{\mu^2}{6\pi^2\epsilon^3}
\int_0^\infty dx x^2\exp(-x^2/2),
\ee
and then using (\ref{iiii}), we obtain (\ref{tildeEE}).

{\bf The photon number}. Combining (\ref{lLKS}),   (\ref{Nph}), and (\ref{SF}),
we arrive at 
\be
\label{st3}
\tilde{\cal N}=\int
\frac{d^3k}{(2\pi)^3}\frac{\BD{J}_\text{mat}(\BD{k})\cdot\BD{J}^*_\text{mat}(\BD{k})}{2\OM{k}^3}.
\ee
For (\ref{JmatS}), we have
\be
\tilde{\cal N}= \frac{\mu^2}{6\pi^2\epsilon^2}
\int_0^\infty dx  x\exp(-x^2/2),
\ee
and then using (\ref{iiii}), we obtain (\ref{tildeNN}).

\section{Integrals for expanding  problem}
\label{NONEQ_app}

{\bf The vector field}. Combining (\ref{lLKS}), (\ref{VVV}), and (\ref{SF}),
we arrive at
\be
\label{ne1}
\lara{\psi(t)|\BD{A}(0,\BD{r})|\psi(t)}=
\lara{\tilde{\psi}(t)|\BD{A}(0,\BD{r})|\tilde{\psi}(t)}-
\int \frac{d^3k}{(2\pi)^3}\BB{\frac{\BD{J}_\text{mat}(\BD{k})}{2\OM{k}^2}\exp(-\ii\OM{k}t+\ii\BD{k}\cdot\BD{r})+\cc}.
\ee
For (\ref{JmatS}), we find using  (\ref{dOmega}) that
\begin{align}
&\lara{\psi(t)|\BD{A}(0,\BD{r})|\psi(t)}=
\lara{\tilde{\psi}(t)|\BD{A}(0,\BD{r})|\tilde{\psi}(t)}
+ 
(\BD{\mu}\times\bnabla)\B{\frac{I_1}{2\pi^2 r}},\\
&I_1=\int_0^\infty dx \frac{\exp(-x^2/4)}{2x}
\BB{\sin\B{x\frac{r-t}{\epsilon}}+\sin\B{x\frac{r+t}{\epsilon}}}.
\end{align}
Proceeding in the same way as with the integral $I$ in Appendix \ref{STAT_app}, we obtain
\be
I_1=\frac{\pi}{4} \BB{\erf\B{\frac{r-t}{\epsilon}}+ \erf\B{\frac{r+t}{\epsilon}}}.
\ee
Using this result and (\ref{AvecS}), we obtain   (\ref{meanA}).

{\bf The free-field energy}. Combining (\ref{lLKS}),  (\ref{dcwihu}), and
(\ref{SF}), we arrive at
\be
\label{ne2}
\lara{\psi(t)|H_0|\psi(t)}=-2\tilde{\cal E}
-\int\frac{d^3k}{(2\pi)^3}\frac{\cos(\OM{k}t)}{\OM{k}^2}\BD{J}_\text{mat}(\BD{k})\cdot\BD{J}^*_\text{mat}(\BD{k}).
\ee
For (\ref{JmatS}), we have
\begin{align}
&\lara{\psi(t)|H_0|\psi(t)}=-2\tilde{\cal E}-\frac{\mu^2}{3\pi^2\epsilon^3}I_2,\\
&I_2=\int_0^\infty dx x^2 \exp(-x^2/2)\cos\B{x\frac{t}{\epsilon}}.
\end{align}
The integral $I_2$  is obtained from formula 3.952.4 of \cite{Ryzhik},
which yields 
\be
I_2=\sqrt{\frac{\pi}{2}}\BB{1-\B{\frac{t}{\epsilon}}^2}\exp\B{-\frac{t^2}{2\epsilon^2}}.
\ee
Using this result and (\ref{tildeEE}), we obtain   (\ref{dmkl}).

{\bf The energy fluctuations}. Combining (\ref{lLKS}),   (\ref{hnn}), and (\ref{SF}), we arrive at
\be
\label{ne3}
h_n=\int \frac{d^3k}{(2\pi)^3}\frac{\BD{J}_\text{mat}(\BD{k})\cdot\BD{J}^*_\text{mat}(\BD{k})}{2\OM{k}^{3-n}}.
\ee
For (\ref{JmatS}), we have
\be
h_n=\frac{\mu^2}{6\pi^2 \epsilon^{n+2}}
\int_0^\infty dx x^{n+1}\exp(-x^2/2),
\ee
and then using (\ref{iiii}), we obtain (\ref{wdfd3}). 

{\bf The number of photons}. Combining   (\ref{lLKS}),
(\ref{Phhhhhh}) with  and (\ref{SF}), we arrive at 
\be
\label{ne4}
{\cal N}(t)=2\tilde{\cal N} - 
 \int \frac{d^3k}{(2\pi)^3}\frac{\cos(\OM{k}t)}{\OM{k}^3}\BD{J}_\text{mat}(\BD{k})\cdot\BD{J}^*_\text{mat}(\BD{k}).
\ee
For (\ref{JmatS}), we have
\begin{align}
&{\cal N}(t)=2\tilde{\cal N}-\frac{\mu^2}{3\pi^2\epsilon^2}I_3,\\
&I_3=\int_0^\infty dx x\exp(-x^2/2)\cos\B{x\frac{t}{\epsilon}}. 
\end{align}
The above integral is obtained from  formula 3.953.4 of \cite{Ryzhik}, which 
after standard simplifications can be written in the 
following compact form
\be
I_3=1-2\frac{t}{\sqrt{2}\epsilon}\daw\B{\frac{t}{\sqrt{2}\epsilon}}.
\ee
Using this result and (\ref{tildeNN}), we obtain (\ref{Ntdaw}).

\section{Derivations of (\ref{daw1}) and (\ref{daw2})}
\label{Dawson}

Equation (\ref{daw1}) is trivially obtained via the Maclaurin expansion. 
Equation (\ref{daw2}) can be obtained by first writing the differential 
equation satisfied by the Dawson function,
\be
\frac{d}{dx}\daw(x)=1-2x\daw(x),
\ee
and then  solving it by substituting the series ansatz 
\be
\daw(x)=\sum_{n=0}^\infty  \frac{a_n}{x^{2n+1}},
\ee
where the coefficients $a_n$ are to be determined.
Note that the structure of this ansatz is chosen 
to be consistent with $\daw(-x)=-\daw(x)$.


%

\end{document}